\documentclass[twocolumn,aps,pra,showpacs]{revtex4}

\usepackage{epsfig}
\usepackage{graphicx}
\usepackage{amssymb}

\begin{document}


\title{Interaction-Free All-Optical Switching via Quantum-Zeno Effect}


\author{Yu-Ping Huang}
\affiliation{Center for Photonic Communication and Computing, EECS Department\\
Northwestern University, 2145 Sheridan Road, Evanston, IL 60208-3118}
\author{Joseph B. Altepeter}
\affiliation{Center for Photonic Communication and Computing, EECS Department\\
Northwestern University, 2145 Sheridan Road, Evanston, IL 60208-3118}
\author{Prem Kumar}
\affiliation{Center for Photonic Communication and Computing, EECS Department\\
Northwestern University, 2145 Sheridan Road, Evanston, IL 60208-3118}

\date{\today}

\begin{abstract}
We propose a novel interaction-free scheme for all-optical switching which does not rely on the physical coupling between signal and control waves. The interaction-free nature of the scheme allows it to overcome the fundamental photon-loss limit imposed by the signal-pump coupling. The same phenomenon protects photonic-signal states from decoherence, making devices based on this scheme suitable for quantum applications. Focusing on $\chi^{(2)}$
waveguides, we provide device designs for traveling-wave and Fabry-Perot switches. In both designs, the performance is optimal when the signal switching is induced by coherent dynamical evolution. In contrast, when the switching is induced by a rapid dissipation channel, it is less efficient.
\end{abstract}
\pacs{42.65.Wi, 42.50.Nn, 42.79.Ta, 03.67.-a} 

\maketitle
\section{introduction}

All-optical networks hold the promise of fast transmission speed and ultra-high data capacity owing to the
non-interacting nature and large bandwidth of photonic signals. A
necessary resource for such networks is all-optical switching,
in which the state of a signal wave is switched in the presence of a control wave (pump). This is routinely achieved using direct coupling between the two
waves in nonlinear materials, such as semiconductors
\cite{TOAD1993}, $\chi^{(3)}$ fibers
\cite{BulVei93,BloDijHoe97,ShaFioKum02,GeuKleKel05}, and cascaded
$\chi^{(2)}$ waveguides
\cite{IroAitArn93,AsoYokIto97,KanItoAso99,KanKumPar01,LeeYuEom06}.
A fundamental limit in these devices is due to the photon loss
resulting from signal-pump coupling. When extended to the quantum
domain, this photon loss additionally causes the quantum states of
photonic signals to collapse (i.e., decohere). To overcome this difficulty, we propose ``interaction-free''
switching that eliminates direct signal-pump coupling.

The concept of ``interaction-free'' phenomenon was first proposed
as ``measurement without touching'' \cite{EliVai93,KwiWinZei96}
and has recently been extended to quantum logic gates
\cite{Azuma03,ZenoGate04,HuaMoo08,Zeno-Semiconductor} and
counterfactual quantum computation \cite{Hosten2006}. In
our application to all-optical switching, we assume the signal wave is initially in state, say,
$|0\rangle$. If un-disrupted, it will coherently evolve
into an orthogonal output state $|1\rangle$, for example, after
passing through a nonlinear waveguide. To induce switching, we apply a pump wave which couples $|1\rangle$ to an ancillary state
$|2\rangle$. This coupling provides a source of disruption to the
$|0\rangle \leftrightarrow |1\rangle$ evolution, resulting in the
signal being frozen in state $|0\rangle$. Thus the signal output
will be switched between $|0\rangle$ and $|1\rangle$ depending on the presence or absence of the pump.  Note that the signal and pump
do not directly interact during operation of the device, as (ideally) the signal remains in $|0\rangle$ when the pump is on. Therefore, in the ideal case, any photon loss from signal-pump coupling is eliminated. Moreover, the quantum states of both the signal and the pump are protected from decoherence.

Each of the above schemes exploits the quantum Zeno effect \cite{zeno1977,zeno1990}, whereby unitary evolution is suppressed via quantum decoherence (i.e.,
quantum measurement)
\cite{zeno_from_decoherence}.  Physically, the decoherence can be induced via either incoherent and irreversible
coupling to many unknown quantum modes (e.g., spontaneous emission into a large number of electromagnetic modes  \cite{Coo88}), or coherent and reversible coupling to a known quantum mode (e.g., polarization decoherence via a polarization-dependent time delay \cite{dfs}).  In the context of quantum measurement, these two processes are mathematically identical. In the context of interaction-free switching, however, we will show that the ``incoherent quantum
Zeno'' (IQZ) effect and the ``coherent quantum Zeno'' (CQZ) effect can lead to
very different switching performances.

An IQZ-based interaction-free switch was recently proposed by Jacobs and Franson (hereafter referred to as the ``JF switch''), which is composed of a microring embedded in an atomic vapor \cite{JacFra09}. We have recently designed another type of CQZ-based, interaction-free switch which is based on the coherent dynamical effect of Autler-Townes
splitting \cite{HuangKumar10}. The latter switch is composed of a
$\chi^{(2)}$ microdisk coupled to two fibers (or waveguides). A
third kind of switch has been demonstrated which relies on
the absorption or scattering of pump photons to change the optical
properties of resonators \cite{VanIbrRit02,IbrCaoKim03,TapLaiLan02,
AlmBarPan04,Switch-Photonic-crytal-2005,Switching-PC2009,Switching-Photonic-Crystal-2005-2}.
Although implemented without direct signal-pump coupling,
switches of this kind are not lossless and involve significant
pump dissipation. They are thus not considered to be
``interaction-free''.

In our switch design, both the $|0\rangle$ and $|1\rangle$ states are assumed to be lossless (a necessary condition for high-fidelity switching).
The ancillary state $|2\rangle$, on the other hand, can
be lossy.  Depending on its lifetime, the interaction-free
disruption will be induced by either the coherent or incoherent
quantum Zeno effects.  First, if $|2\rangle$ is short lived,
then the $|1\rangle\rightarrow|2\rangle$ coupling effectively
opens a dissipation channel for $|1\rangle$. The inhibition
of the $|0\rangle \leftrightarrow |1\rangle$ transition is
therefore a consequence of the IQZ effect. In the second
case, that of a long-lived $|2\rangle$ sate, there is no
dissipation involved. Instead, the $|0\rangle \leftrightarrow
|1\rangle$ transition is suppressed by an Autler-Townes--based CQZ
effect \cite{AutTow55}, similar to the electromagnetic-induced
transparency \cite{EIT-Rev05}. The former IQZ effect
corresponds to \emph{broadening} of the $|1\rangle$ state,
whereas the latter CQZ effect corresponds to \emph{shifting} of
the same state. The two processes are equally effective for
suppressing the $|0\rangle\leftrightarrow|1\rangle$ transition,
but lead to very different pump-power requirements for similar
switching setups. To see why, we consider the toy ``$\Lambda$''
model described above with Rabi frequencies of $\Omega_s$
and $\Omega_p$ for the $|0\rangle\leftrightarrow|1\rangle$ and
$|1\rangle\leftrightarrow|2\rangle$ transitions, respectively. The
system evolution time $t$ is set to be $\pi/\Omega_s$, such that the
signal evolves to $|1\rangle$ with pump off. The $|2\rangle$
state is assumed to decay at rate $\gamma$. Note that
a necessary condition is $|\Omega_p|\gg |\Omega_s|$. In the IQZ regime with $\gamma\gg\Omega_p$, one finds via adiabatic elimination that the output power fraction of the signal in $|1\rangle$ with pump on is $\gamma t |\Omega_s|^2/|\Omega_p|^2$.  In contrast, in the CQZ regime with $\gamma\ll \Omega_p$, the power fraction in $|1\rangle$ is $\le |\Omega_s|^2/|\Omega_p|^2$. As $\gamma t \gg |\Omega_p| t\gg \pi$, the fraction of the signal evolving into $|1\rangle$,
even with the pump on, is much higher in the IQZ regime than in
the CQZ regime. This strongly suggests that the CQZ effect is more
efficient in suppressing the $|0\rangle\leftrightarrow|1\rangle$
transition, which will lead to better switching performance in
the CQZ regime. Physically, this is because in the presence of a
strong dissipation for $|2\rangle$, the $|1\rangle\leftrightarrow
|2\rangle$ coupling is adversely suppressed by the IQZ effect. As
a result, for a given pump, the $|0\rangle\rightarrow |1\rangle$
transition is in fact less disrupted in the IQZ regime.

In order to perform a systematic study of the interaction-free switching, in this paper we consider systems composed of $\chi^{(2)}$ waveguides. Two
designs are provided. The first is a traveling-wave design composed of a single-pass waveguide. With no pump, the signal
undergoes a coherent second-harmonic generation (SHG) process
in the waveguide, resulting in frequency doubling or $\pi$-phase
shifting. In this design, the fundamental (signal input) and the
second-harmonic (SH) wave correspond to $|0\rangle$ and
$|1\rangle$, respectively, in the above prototypical model. To induce switching, a pump wave suppresses the SHG by coupling the SH state to an ancillary state ($|2\rangle$). This coupling
can be either sum-frequency generation (SFG) or
difference-frequency generation (DFG), but for concreteness, in this traveling-wave design we focus on DFG. Using both the IQZ and CQZ effects, we are able to construct frequency, polarization, and spatial-mode switches. To achieve similar performances, a much stronger
pump is required in the IQZ regime than in the CQZ regime. All
switches turn out to work well for continuous-wave inputs. For the realistic case of non-flat pulses, however, only the frequency switch is potentially practical. This is because both the polarization and spatial-mode switches are
subject to significant pulse distortion, which fundamentally
arises from the nonlinear nature of the SHG process. Such distortion would severely limit the cascadability of this type of switches, making them not very useful for potential practical applications.

To overcome pulse distortion, our second design utilizes a Fabry-Perot cavity, specifically, a $\chi^{(2)}$ waveguide coated with reflective layers on the two end faces. The cavity is designed to be resonant for both the signal and pump waves. In the absence of the pump, the signal wave, say, applied from the left, is resonantly coupled with the
cavity mode. Eventually, it exits from the right end after a time delay. To induce switching, we apply a pump wave to the cavity, also through the
left-end layer. In the cavity, the signal and pump waves undergo SFG (or
DFG). The SFG dynamics shifts the cavity out of resonance, resulting in the signal being reflected from the cavity. Note here that the SFG only ``potentially'' happens, as ideally the signal would never enter the cavity. Moreover, by
applying the pump pulse slightly ahead of the signal pulse, the pump will pass through the cavity unaffected (as there is ideally no signal field in the cavity). Compared to the traveling-wave design, in this case the quasi-linear cavity-coupling process would overcome pulse distortion, provided that the cavity passband is wider than the Fourier spectrum of the pulses. Again, in this design, better switching performance is obtained in the CQZ regime than in the IQZ regime. We note that, in the IQZ regime, our cavity design is physically equivalent
to the JF switch \cite{JacFra09}. Both are implemented via avoided
two-photon absorption. Physically, the JF device is composed of a microring evanescently coupled to an atomic vapor. In contrast, our device utilizes an all solid-state design. In the CQZ regime, the present Fabry-Perot switch can be mapped onto our recently proposed microdisk switch \cite{HuangKumar10}. The goal of this paper is to systematically study the
IQZ and CQZ effects in interaction-free switching by varying the system dissipation. We present the traveling-wave and Fabry-Perot designs in sections \ref{ss} and \ref{FPC}, respectively, before presenting a brief conclusion in section \ref{dc}.

\section{Traveling-wave Switch}
\label{ss}
In this section we present the traveling-wave design for interaction-free
all-optical switching. In section \ref{ss-setup} we describe the schematic
setups for three types of devices. In section \ref{wd}, we analytically solve the
system dynamics in the CQZ and IQZ regimes. Using these results, in section \ref{pa} we analyze the switching performance for both continuous-wave and pulsed inputs.

\subsection{Setup}
\label{ss-setup}
\begin{figure}
\centering \epsfig{figure=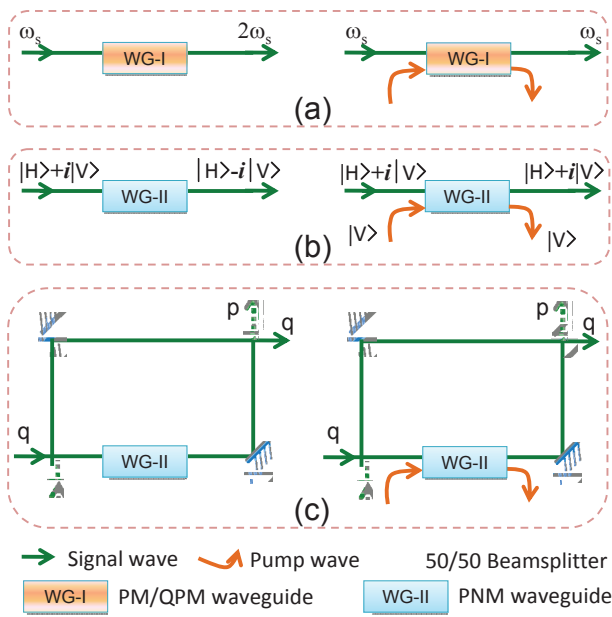, width=8.5cm}
\caption{(Color online) Schematic setups for frequency switching (a), polarization switching (b), and spatial-mode switching (c). \label{fig1}}
\end{figure}

The schematic setups for our frequency, polarization, and spatial-mode switching devices are shown in Figs.~\ref{fig1} (a), (b), and
(c), respectively. The frequency switch transforms a
signal wave originally at angular frequency $\omega_s$ into an identical
signal at angular frequency $\omega_s$ or $2\omega_s$, depending on the
presence or absence of the control wave. To achieve this, the
waveguide is designed to be phase-matching (PM) or quasi-phase-matching (QPM) for SHG of the signal wave. Phase-matching can be achieved in some birefringent crystals by angle tuning \cite{BorWol75} or temperature tuning \cite{HobWar66}, whereas quasi-phase-matching is more flexibly realized in periodically-poled materials whose crystal-axis orientation is periodically inverted along the wave-propagation direction \cite{ArmBloDuc62,YamNadSai93}. In the absence of the control wave, the signal wave undergoes a complete SHG cycle as it travels through the waveguide, where its power is monotonically transferred to the harmonic wave (at angular frequency $2\omega_s$). In the presence of the control wave the SHG process is suppressed, resulting in the majority of the signal power remaining at angular frequency $\omega_s$. In effect, the presence of the control wave switches the output signal frequency from $2\omega_s$ to $\omega_s$.

Analogously, the polarization switch (see Fig.~\ref{fig1}(b))
performs a polarization rotation (while leaving the signal
frequency unchanged) that is conditional on the presence
or absence of the pump wave. This type of switch employs
phase-mismatched (PNM) waveguides to avoid monotonous
transfer of power from the signal to the second-harmonic wave. Using a uniaxial crystal, the waveguide is designed such that one of the two orthogonal signal polarizations experiences a $\pi$ phase shift during
the SHG process. This can be achieved, for example, by using a periodically-poled (PP) uniaxial waveguide in which the ordinary-polarized light is close to QPM, whereas the extraordinary-polarized light is
far from phase matching. As above, the presence of the control
wave suppresses the SHG process and the $\pi$ phase shift
associated with it.  As an example, consider a waveguide
designed such that the polarization state $|V\rangle$ gains an
SHG-induced $\pi$ phase shift, while the orthogonal polarization
state $|H\rangle$ experiences no SHG and hence no phase shift.  This
waveguide will rotate an input signal whose polarization state
is $|+\rangle\equiv\frac{1}{\sqrt{2}} (|H\rangle+i|V\rangle)$
to an output signal with polarization state $|-\rangle\equiv\frac{1}{\sqrt{2}}
(|H\rangle-i|V\rangle)$.  When the pump wave is applied, the
SHG process is suppressed, leaving the input signal's polarization
state $|+\rangle$ unchanged.  To implement this switch in
isotropic crystals, the waveguide must be designed such that
both polarizations experience an identical $\pi$-phase shift in
the absence of the pump beam. An input signal $|+\rangle$ will
transform to the state $-|+\rangle$, i.e., unchanged except for a
global $\pi$ phase shift.  The switching operation is induced by a vertically-polarized pump of appropriate intensity which suppresses
the SHG process for vertically-polarized inputs, thus eliminating the
SHG-induced $\pi$ phase shift.  This type of operation is feasible
in type-I $\chi^{(2)}$ crystals because the diagonal nonlinear coefficient
$d_{33}$ is usually much larger than the off-diagonal ones.  For a
control wave of a particular intensity, one can strongly disturb the
$|V\rangle$-polarized harmonic wave (up-converted from the $|V\rangle$-polarized signal wave), while leaving the $|H\rangle$-polarized harmonic wave nearly undisturbed. (An analogous design is also possible for biaxial crystals.)

By applying the same type of $\pi$-phase suppression to an input signal in a waveguide inside one arm of an interferometer, we can realize an
analogous spatial-mode switching device.  The schematic setup in Fig.~\ref{fig1}(c) shows a Mach-Zehnder interferometer containing a waveguide in the lower arm.  This waveguide is designed such that
the SHG process induces a $\pi$ phase shift at the signal wavelength. In the absence of a control wave, the relative phase between the two arms of the interferometer can be adjusted such that a signal wave entering the ``q''
input port of the interferometer will exit from the ``q''
output port of the interferometer. In the presence of the control
wave, the SHG process is suppressed along with the associated phase shift. As a result, the signal wave will exit from the ``p'' output
port of the interferometer.  This setup can be
modified to incorporate two symmetric waveguides with opposite-sign
phase-mismatching coefficients, one in each arm of the interferometer, such that---in the absence of
the control wave---phase shifts of $\pi/2$ and $-\pi/2$ will be
applied to the signal wave in each arm.

\begin{figure}
\centering \epsfig{figure=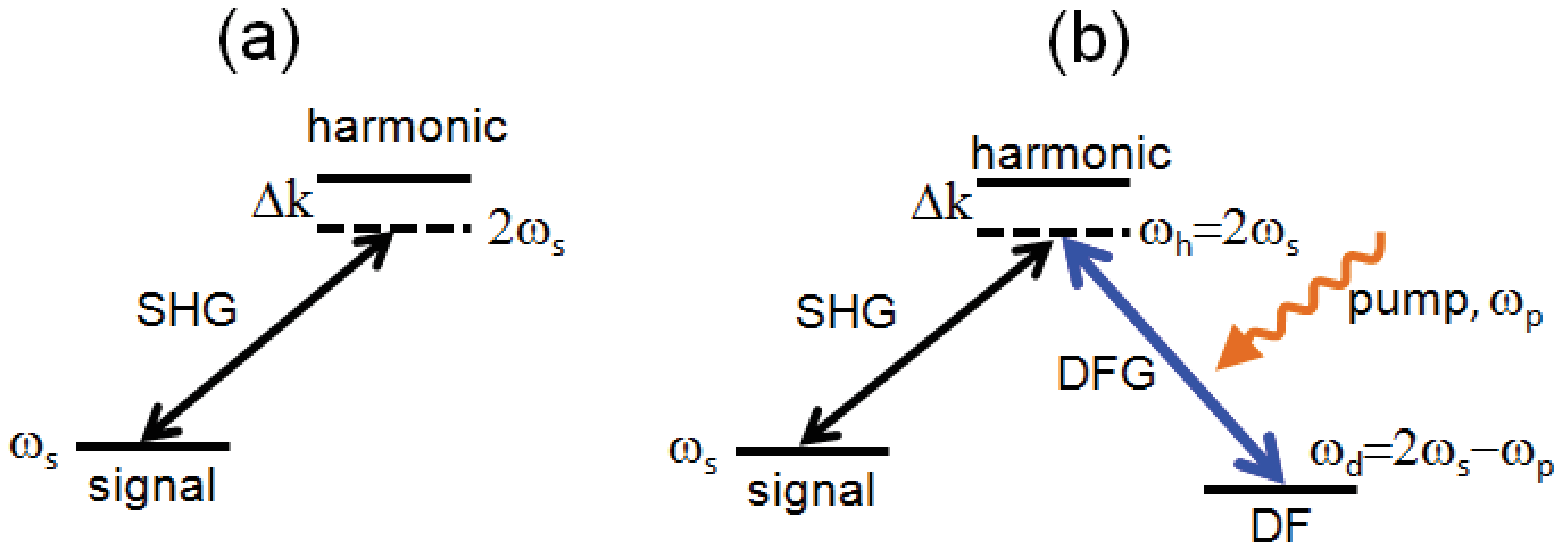, width=8.5cm}
\caption{(Color online)
The level scheme for DFG-based switching, showing the cases for pump on (a)
and pump off (b).
\label{fig2}}
\end{figure}

Fundamentally, the switches described here rely
on second-order nonlinearities to enable the
optical control.  More specifically, they rely on the
suppression of certain second-order nonlinear optical processes
in the presence of a pump wave. There are two ways to
achieve this: sum-frequency generation (SFG) and difference-frequency
generation (DFG).  When using SFG, the harmonic wave (with angular frequency $\omega_h=2\omega_s$) is coupled to the pump wave (with angular frequency
$\omega_p$) and up-converted into a sum-frequency
(SF) wave of angular frequency $\omega_{\ell}=\omega_h+\omega_p$. When
using DFG, the harmonic and pump waves generate a difference-frequency
(DF) wave of angular frequency $\omega_d=\omega_h-\omega_p$ (or equivalently, $\omega_d=\omega_p-\omega_h$), as shown in Fig.~\ref{fig2}. In both cases, the harmonic level (following atomic-physics terminology) will be disturbed and, provided
that the coupling is sufficiently strong, the SHG process will
be suppressed due to either the IQZ or the CQZ effect. If the signal wave is in a single-mode of the waveguide, the SF wave will be well guided; such may not be the case for the DF wave, however. For example, when using a $1565$-nm signal wave and a $1535$-nm pump wave,
the resulting DF wave will be in the terahertz (THz) regime. For
typical waveguides, such as annealed proton-exchanged waveguides in PPLN with a small circumference ($\sim$\,100 $\mu$m), such a THz wave
will be strongly diffracted, leading to energy dissipation
from the system. This provides a natural dissipation channel
for the IQZ effect via the purely geometrical effect
of \emph{diffraction}. (If desired, this diffraction could be overcome by applying reflective coatings to the waveguide side surfaces.) In contrast, for the SF wave the potential dissipation must rely solely on material absorption. For concreteness, in the following we will focus on the DFG scheme as shown in Fig.~\ref{fig2}. Extending to the SFG case is straightforward.

In all of the optical switches described above, the pump wave
does not directly interact with the signal wave. This is because the signal's second-harmonic level---the level coupled to the pump wave---never gets populated when under the suppressive influence of the pump. Again, this interaction-free feature eliminates photon loss and protects against decoherence.

\subsection{Wave Dynamics}
\label{wd}

To model the system dynamics, we denote the positive-frequency
electric fields for the signal, harmonic, pump, and DF waves as $E_s, E_h, E_p$, and $E_d$, respectively. We assume all waves travel
in the $\hat{z}$ direction and are linearly
$\hat{x}$-polarized. Note whereas only the type-I dynamics are considered here, generalization to the type-II case is straightforward.

To proceed, we define the following dimensionless variables,
\begin{equation}
\label{eq-define}
    A_j=i\sqrt{\frac{\epsilon_0 V n_j}{2\hbar\omega_j}} E_j e^{i\beta_j z-i\omega_j t},
\end{equation}
with $j=s,h,p,d$. Here, $V$ is the quantization volume, $\epsilon_0$ is the permittivity of free space, $k_j$ is the wave-vector magnitude, and $n_j$ is the refractive index. Defined as such, the quantity $|A_j|^2$ gives the average number of photons inside the volume $V$. Further, we introduce
the effective Rabi-frequencies in the space domain as
\begin{eqnarray}
\label{eqn2}
    \Omega_1 &=& 4 d_{\mathrm{eff}}\sqrt{\frac{\hbar\omega_s^3}{n_s^2 n_h \epsilon_0 c^2 V}}, \\
\label{eqn3}
    \Omega_2 &=& 2 d_{\mathrm{eff}} \sqrt{\frac{2\hbar\omega_h \omega_p \omega_{d}}{n_p n_d n_h \epsilon_0 c^2 V}},
\end{eqnarray}
where $d_{\mathrm{eff}}$ is the effective second-order nonlinear coefficient.

In this paper, we assume a matched group velocity $v_g$ for all the waves and neglect any group-velocity dispersion. For switching, we consider a phase-matched DFG process. These two conditions lead to optimal switching performance. Furthermore, for simplicity we assume all waves to be lossless except for the DF wave. The coupled-wave equations in the moving frame $z\rightarrow z-t/v_g$ are
\begin{eqnarray}
\centering
\label{eqn5}
    \frac{\partial A_{s}}{\partial z} &=&-\Omega_1 e^{-i\Delta k z}A_{h} A^\ast_{s}, \\
    \frac{\partial A_{h}}{\partial z} &=& \frac{\Omega_1}{2}  e^{i\Delta k z} A^2_{s}
     +\Omega_2  A_{p} A_{d}, ~\\
         \frac{\partial A_{p}}{\partial z}&=&-\Omega_2 A^\ast_d A_{h}, \\
\label{eqn6}
      \frac{\partial A_{d}}{\partial z}&=&-\Omega_2 A^\ast_p  A_{h}-\gamma A_{d}.
\end{eqnarray}
Here, $\Delta k=2k_s-k_h$ is the phase-mismatch per unit length and $2\gamma$ is the decay rate of the DF wave.

\subsubsection{Case I: Pump off}
\begin{figure}
\centering \epsfig{figure=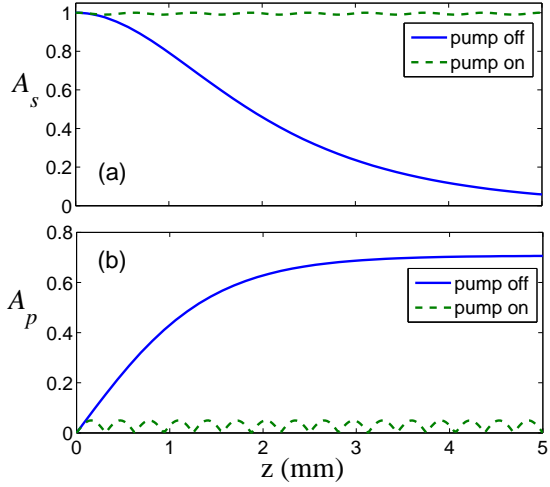, width=8.0cm}
\caption{(Color online) Wave dynamics in a PM waveguide with the pump off and on,
for (a) the signal wave and (b) the harmonic wave. Parameters are chosen as follows: $\Omega_\mathrm{eff}=10 ~\mathrm{mm}^{-1}$, $\Omega_1=1 ~ \mathrm{mm}^{-1}$,  $A_s(0)=1$, $\Delta k=0 ~\mathrm{mm}^{-1}$ and $\kappa=0 ~\mathrm{mm}^{-1}$.
\label{fig3}}
\end{figure}

In the absence of the pump wave, i.e., $A_p(0)=0$, the equations
of motion are reduced to the well-known SHG equations
\cite{ArmBloDuc62,LiKum94}. For our purposes, there are three dynamical regimes of interest: a) Perfect phase-matching with $\Delta k=0$; b) small PNM with $\Delta k\ll \Omega_1 |A_s|$ ($\Delta k\neq 0$); and c) large PNM with $\Delta k\gg \Omega_1 |A_s|$. Defining the field amplitude and phase as $A_j=\mu_j e^{i\phi_j}$, for $\Delta k=0$, the SHG dynamics have an analytical solution \cite{LiKum94}:
\begin{eqnarray}
\label{eqn19}
    \mu_s(z) &=& {\mu^0_s}~ \mathrm{sech}\left(\frac{{\mu^0_s}\Omega_1 z}{\sqrt{2}}\right), \\
    \mu_h(z) &=& \frac{{\mu^0_s}}{\sqrt{2}}~\mathrm{tanh}\left(\frac{{\mu^0_s}\Omega_1 z}{\sqrt{2}}\right),  \\
    \phi_s(z) &=& \phi_s(0), \\
    \phi_h(z) &=& 2\phi_s(0).
\end{eqnarray}
where $\mu^0_s\equiv\mu_s(0)$. Thus the system undergoes a monotonous up-conversion process, transferring all power from $\omega_s$ to $2\omega_s$. An example of this PM dynamics is shown in Fig.~\ref{fig3}.

For a PNM waveguide with $\Delta k\neq 0$, a general analytical solution was also derived \cite{ArmBloDuc62}, where the amplitudes are given by
\begin{eqnarray}
\label{eqn7a}
   & & \!\!\!\!\!\!\mu_s(z)={\mu^0_s} \sqrt{1-\frac{1}{\chi^2} \mathrm{JacobiSN}^2 \left[\frac{\chi {\mu^0_s}\Omega_1 z}{\sqrt{2}}, \frac{1}{\chi^4}\right]}, ~~~~~\\
\label{eqn8a}
   & &\!\!\!\!\!\!\mu_h(z)={\mu^0_s} \frac{1}{\sqrt{2}\chi} \left| \mathrm{JacobiSN} \left[\frac{\chi {\mu^0_s} \Omega_1 z}{\sqrt{2}}, \frac{1}{\chi^4}\right]\right|,
\end{eqnarray}
with
\begin{equation}
    \chi=\frac{\sqrt{2}\,\Delta k}{4\Omega_1 \, {\mu^0_s}}+\sqrt{1+\frac{\Delta k^2}{8\Omega^2_1 \, \mu^2_s(0)}}.
\end{equation}
The solutions (\ref{eqn7a}) and (\ref{eqn8a}) are oscillatory functions with a period
\begin{equation}
\label{eqn21}
    z_0=\frac{2 \sqrt{2}}{\chi {\mu^0_s} \Omega_1} \mathrm{EllipticK}\left(\frac{1}{\chi^4}\right).
\end{equation}

\begin{figure}
\centering \epsfig{figure=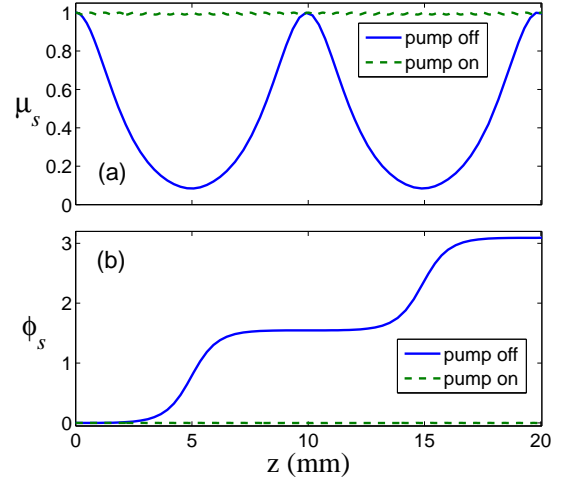, width=8.0cm} \caption{(Color online)
Wave dynamics for small PNM with the pump off and on. In (a) we plot the amplitude and in (b) the phase of the signal wave. Parameters are: $\Omega_\mathrm{eff}=10~ \mathrm{mm}^{-1}$, $\Omega_1=1 ~\mathrm{mm}^{-1}$, $A_s(0)=1$, $\Delta
k=0.01 ~\mathrm{mm}^{-1}$ and $\gamma=0$.
\label{fig4}}
\end{figure}

\begin{figure}
\centering \epsfig{figure=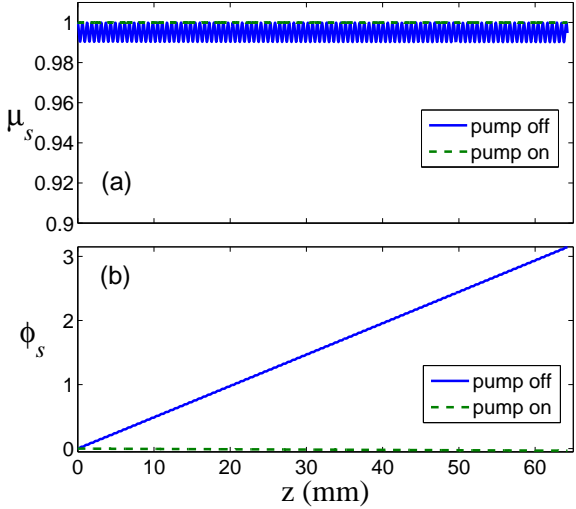, width=8.5cm} \caption{(Color online)
Wave dynamics for large PNM with the pump off and on. In (a) and (b) we show the amplitude and phase of the signal wave, respectively. Parameters are: $\Omega_\mathrm{eff}=100~ \mathrm{mm}^{-1}$, $\Omega_1=1 ~\mathrm{mm}^{-1}$, $A_s(0)=1$, $\Delta
k=10 ~\mathrm{mm}^{-1}$ and $\gamma=0$.
\label{fig5}}
\end{figure}

In the small PNM regime with $\Delta k<\Omega_1
{\mu^0_s}$, the signal and pump waves undergo a complete
power-conversion process, whereas in the opposite regime of
$\Delta k>\Omega_1 {\mu^0_s}$, the majority of power is trapped
in the signal wave. In the limit $\Delta k\gg \Omega_1
{\mu^0_s}$, we have
\begin{eqnarray}
\label{eqn14}
   \mu_s(z)&=&{\mu^0_s}-\frac{\Omega_1^2 (\mu^0_s)^3}{\Delta k^2} \sin^2 \left(\frac{\Delta k z}{2}\right), \\
   \mu_h(z)&=& \frac{(\mu^0_s(0))^2 \Omega_1}{\Delta k} \left|\sin \left(\frac{\Delta k z}{2}\right)\right|.
\end{eqnarray}
In this case, only a small fraction ($\frac{2\mu^2_s(0)
\Omega_1^2}{ \Delta k^2} \ll 1$) of the signal power
oscillates between the fundamental and SH levels. As an example, in Figs.~\ref{fig4}(a) and~\ref{fig5}(a) we plot the amplitude dynamics of the fundamental wave for the cases of small and large PNM, respectively.

The phase dynamics, on the other hand, yield a complicated and nonintuitive expression. Nonetheless, for $\Delta k \lesssim 0.01 \Omega_1 {\mu^0_s}$, the periods for the fundamental wave to restore its power while gaining a phase shift of $\pi/2$ and $\pi$ are approximately $z_0$ and $2 z_0$, respectively. In the large PNM limit ($\Delta k\gg \Omega_1 {\mu^0_s}$), the phase shift increases linearly with $z$, giving
\begin{equation}
\label{eqn10}
    \phi_s(z)=\frac{\Omega_1^2 \mu^2_s(0)}{2\Delta k} z.
\end{equation}
This result also shows a linear dependence of the phase shift on the intensity of the signal wave. As we will show later, this intensity-dependence constitutes a major obstacle to applying the present traveling-wave switching scheme to time-varying pulses. The phase dynamics in the small and large PNM regimes are shown in Figs.~\ref{fig4}(b) and~\ref{fig5}(b), respectively.

\subsubsection{Case II: Pump on}
In the presence of the pump wave, the system dynamics will be
disturbed. For switching purposes, the dynamical regime of interest is
where the SHG process is strongly suppressed by either the IQZ or the CQZ effect. In either case, a strong pump wave will be required such that $\Omega_2 A_p \gg \Omega_1 A_s$, as we will show later. For successful operation, the harmonic wave must remain unpumped while the pump wave remains undepleted (the ``interaction-free'' regime). Under the undepleted-pump approximation, for which $A_p(z)=A_p(0)$, Eqs.~(\ref{eqn5})--(\ref{eqn6}) are simplified to
\begin{eqnarray}
\centering
\label{eqn7b}
    \frac{\partial A_{s}}{\partial z} &=&-\Omega_1 e^{-i\Delta k z}A_{h} A^\ast_{s}, \\
    \frac{\partial A_{h}}{\partial z} &\approx& \frac{\Omega_1}{2}  e^{i\Delta k z} A^2_{s}+\Omega_{\mathrm{eff}} A_{d}, ~\\
\label{eqn8b}
      \frac{\partial A_{d}}{\partial z}&=&-\Omega_{\mathrm{eff}}  A_{h}-\gamma A_{d},
\end{eqnarray}
where we have introduced
\begin{equation}
    \Omega_{\mathrm{eff}}=\Omega_2 A_p(0),
\end{equation}
and, without loss of generality, assumed $A_p(0)$ to be real.

There are two characteristic dynamical regimes governed by Eqs.~(\ref{eqn7b})-(\ref{eqn8b}). In the first regime, $\gamma\ll
\Omega_{\mathrm{eff}}$, i.e., the loss rate for the DF-wave is
much smaller than the DFG rate. In this case, the SHG
process will be suppressed by the CQZ effect. To obtain
an analytic solution, we introduce the following dressed-state waves that are decoupled in the Hilbert space defined by the DFG hamiltonian:
\begin{equation}
    A_{\pm}=\frac{1}{2} e^{-i\Delta k z}(A_h\pm iA_d).
\end{equation}
The wave equations (\ref{eqn7b})--(\ref{eqn8b}) can then be rewritten as
\begin{eqnarray}
\centering
\label{eqn9}
    \frac{\partial A_{s}}{\partial z} &=&-\Omega_1 (A_{+}+A_{-}) A^\ast_{s}, \\
  \label{eqn12}
    \frac{\partial A_{\pm}}{\partial z} &=& \frac{\Omega_1}{4}  A^2_{s}-i(\Delta k\pm \Omega_{\mathrm{eff}}) A_{\pm}.
\end{eqnarray}
These equations clearly show that the effect of a strong pump coupling is to split the harmonic level into two dressed
levels detuned from the original SHG resonance by $\pm
\Omega_\mathrm{eff}$. Given that $\Omega_\mathrm{eff}\gg \Omega_1 |A_s|$, the SHG dynamics are shifted off-resonance and are thus suppressed by the CQZ effect. In this picture, our system is similar to an EIT system.

To obtain analytical solutions to Eqs.~(\ref{eqn9})
and~(\ref{eqn12}), we follow the standard method of separating
the fastest "time" scales in order to obtain the adiabatic-elimination results \cite{ScuZub97},
\begin{equation}
\label{eqn17}
    A_\pm (z)\approx \frac{-i\Omega_1 A_s^2(z)}{4(\Delta k\pm\Omega_\mathrm{eff})},
\end{equation}
which are valid in the limit of $|\Delta k\pm\Omega_\mathrm{eff}|\gg \Omega_1 {\mu^0_s}$. Inserting this result into Eq.~(\ref{eqn9}), one arrives at an effective equation of motion for $A_s$, where
\begin{equation}
    \frac{\partial A_{s}}{\partial z} =\frac{i\Omega^2_1}{2\Delta k}
    \mathcal{S}_\mathrm{CQZ}  |A_{s}|^2 A_s
\end{equation}
with the phase-suppression coefficient
\begin{equation}
     \mathcal{S}_\mathrm{CQZ}=\frac{\Delta k^2}{\Delta k^2+\Omega^2_\mathrm{eff}}.
\end{equation}
Solving the above equation to first order gives
\begin{equation}
\label{eqn11}
     A_s(z)=\exp\left(\frac{i\Omega^2_1 (\mu^0_s)^2 z}{2\Delta k}
     \mathcal{S}_\mathrm{CQZ}\right) \mu^0_s.
\end{equation}
This result exhibits a pure phase evolution. In the limit of $\Omega_\mathrm{eff}\ll \Delta k$, the underlying
phase-shift dynamic recovers the result of Eq.~(\ref{eqn10})
since $\mathcal{S}_\mathrm{CQZ}\rightarrow 1$. Thus there is no suppression
effect. In the opposite limit of $\Omega_\mathrm{eff}\gg \Delta k$, in
contrast, the phase shift is suppressed by a factor of
$\mathcal{S}_\mathrm{CQZ}\approx (\Delta k/\Omega_\mathrm{eff})^2\ll 1$
smaller than in the pump-off case. The $1/\Omega^2_\mathrm{eff}$ scaling in
$\mathcal{S}_\mathrm{CQZ}$ clearly suggests a high-efficiency method for controlling the phase shift of the signal wave.

In Eq.~(\ref{eqn11}), the amplitude of $A_s(z)$ is a
constant of motion, resulting from the first-order approximation that was used to keep only the slowest-varying term in Eq.~(\ref{eqn17}) for $A_\pm$. The fast-varying part of $A_\pm$ can be obtained
iteratively by using the result in Eq.~(\ref{eqn11}). By inserting the
resulting expression for $A_\pm$ into Eq.~(\ref{eqn9}), we can
get an effective equation of motion for $A_s$, which is a second-order result. In the parameter
regime of interest ($\Omega_\mathrm{eff}\gg \Delta k$),
the amplitude dynamics are governed by
\begin{equation}
    \frac{\partial \mu_s}{\partial z}\approx-\frac{\Omega^2_1 \sin(\Omega_\mathrm{eff} z)}{2\Omega_\mathrm{eff}} \mu_s^3,
\end{equation}
such that
\begin{equation}
\label{eqn13}
    \mu_s(z)={\mu^0_s}\left(1+\frac{2\Omega_1^2 \mu^2_s(0)}{\Omega_\mathrm{eff}^2} \sin^2(\frac{\Omega_\mathrm{eff} z}{2})\right)^{-1/2}.
\end{equation}
Here, the amplitude of the fundamental wave rapidly oscillates
between its initial value ${\mu^0_s}$ and
${\mu^0_s}\left(1-\frac{\Omega^2_1
\mu^2_s(0)}{\Omega^2_\mathrm{eff}}\right)$. In effect, only a
fraction of the input power ($\frac{2\Omega^2_1 |A_s(0)|^2}{\Omega^2_\mathrm{eff}}\ll 1$) undergoes frequency conversion, while most remains in the fundamental wave. We note that during this
dynamics no dissipation of any kind is involved; rather, the SHG
is suppressed by the coherent CQZ effect. In
principle, all the input power can be restored to the fundamental wave (the
desired signal output), thus implementing a lossless switch. In
practice, however, it may be noticeably difficult to achieve this, as
the frequency-conversion oscillation has a very short period ($\sim
1/\Omega_\mathrm{eff}$).

In Figs.~\ref{fig3}, \ref{fig4}, and~\ref{fig5} we
compare the wave dynamics with the pump off and on in the
regimes of PM (Fig.~\ref{fig3}), small phase-mismatching (Fig.~\ref{fig4}), and large PNM (Fig.~\ref{fig5}),
respectively. In all the dynamical regimes, the amplitude and phase
dynamics are strongly suppressed by the CQZ effect.

Finally, we note that by replacing $\Omega_\mathrm{eff}$ with
$\Delta k$, the leading-order term in the Taylor series of Eq.~(\ref{eqn13}) is the same as Eq.~(\ref{eqn14}). This result
clearly shows that the effect of strong DFG coupling is to shift
the SHG off resonance (i.e., off phase-matching).

In the first dynamical regime considered above, the CQZ effect dominates ($\gamma\ll\Omega_\mathrm{eff}$). In contrast, in the second, opposite, dynamical regime of interest with $\gamma\gg\Omega_\mathrm{eff}$, the
wave dynamics are dominated by the IQZ effect. To analytically
solve for the dynamics in this regime, we adiabatically eliminate
the DF wave in Eqs.~(\ref{eqn7b})--(\ref{eqn8b}), obtaining
\begin{eqnarray}
\centering
\label{eqn15}
    \frac{\partial A_{s}}{\partial z} &=&-\Omega_1 e^{-i\Delta k z}A_{h} A^\ast_{s}, \\
\label{eqn16}
    \frac{\partial A_{h}}{\partial z} &\approx& \frac{\Omega_1}{2}  e^{i\Delta k z} A^2_{s}-\kappa_\mathrm{eff} A_{h}
\end{eqnarray}
with the effective decay rate for the SH level given by
\begin{equation}
\label{eqn18}
    \kappa_\mathrm{eff}=\frac{\Omega_{\mathrm{eff}}^2}{\gamma}.
\end{equation}
Here, the effect of coupling to a lossy DF wave is to open a decay
channel for the SH level. Interestingly, the effective decay rate
$\kappa_\mathrm{eff}$ is inversely proportional to the DF loss
rate $\gamma$. This is a consequence of the IQZ-suppression effect
on the DFG dynamics (not SHG). More specifically, in the presence
of a strong dissipation for the DF wave, the DFG dynamics are
suppressed by the IQZ effect, leading to
$A_d\approx\frac{\Omega^2_{\mathrm{eff}}}{\gamma^2} A_h \ll A_h$.
Thus the DF wave remains nearly unpumped. As a result, the effective
dissipation rate for the SH level is $\gamma A_d/A_h$, giving
Eq.~(\ref{eqn18}).

\begin{figure}
\centering \epsfig{figure=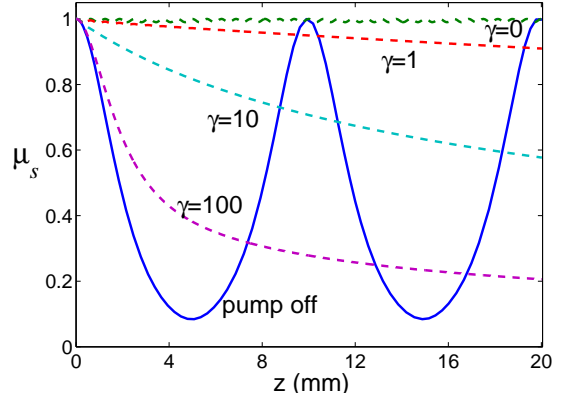, width=8.0cm} \caption{(Color online)
Amplitude dynamics of the signal wave with the pump off (solid line) and pump on (dashed lines). The chosen parameters are: $\Omega_\mathrm{eff}=10 ~\mathrm{mm}^{-1}$,
$\Omega_1=1~ \mathrm{mm}^{-1}$, $A_s(0)=1$, $\Delta k=0.01 ~ \mathrm{mm}^{-1}$. The DF dissipation rates $\gamma$ marked in the figure are in units of $\mathrm{mm}^{-1}$.
\label{fig6}}
\end{figure}

In the IQZ regime, a necessary condition for successful switching
is $\kappa_\mathrm{eff}\gg \Omega_1 \mu_s$ in the case of PM SHG or
small-PNM SHG, or $\kappa_\mathrm{eff}\gg \Omega^2_1
\mu_s^2/\Delta k$ in the case of large-PNM SHG. Given that, the wave
dynamics can be solved by adiabatically eliminating the SH wave and we obtain
\begin{eqnarray}
     A_s &=& A_s(0) \left(1+\frac{\Omega^2_1 \mu^2_s(0)}{\kappa_\mathrm{eff}+i\Delta k}z\right)^{-1/2} \\
            &\approx& A_s(0) \exp\left(-\frac{\Omega^2_1 \mu^2_s(0)
            \gamma}{2\Omega^2_\mathrm{eff}}z\right)
            \exp\left(i\frac{\Omega^2_1 \mu^2_s(0)}{2\Delta k}
            \mathcal{S}_\mathrm{IQZ}  z\right) \nonumber
\end{eqnarray}
with a phase suppression coefficient
\begin{equation}
   \mathcal{S}_\mathrm{IQZ}=\frac{\Delta k^2 \gamma^2}{\Omega^4_\mathrm{eff} }.
\end{equation}
Here, both the amplitude loss and the phase shift of the signal wave
increase with $\gamma$. Thus a better switching performance is
achieved with a weaker loss of the DF wave; in fact, the switch is
optimal for $\gamma=0$ corresponding to the CQZ regime. To show
this, in Fig.~\ref{fig6} we plot the SHG dynamics for various values of
$\gamma$. As shown, the SHG dynamics is most efficiently
suppressed with $\gamma=0$, i.e., in the CQZ regime. As $\gamma$
increases, the suppression effect becomes weaker. Eventually, for
$\gamma\rightarrow \infty$, the SHG dynamics approach the pump-off
result.

Of course, the fact that the SHG-suppression is weaker with a
large $\gamma$ does not imply that the IQZ effect is inapplicable
to all-optical switching. Our conclusion is that for a given pump
intensity, the SHG-suppression effect, and thus the switching
efficiency, is highest when the system is operated in the CQZ regime. In
other words, to achieve a certain switching performance, a much weaker pump
would be required when the switch is operated in the CQZ regime than in the IQZ regime. In terms of photon loss, a switching operation in the CQZ regime yields a fraction of
$\le 2\Omega_1^2 \mu_s^2(0)/\Omega^2_\mathrm{eff}$ compared to
$\Omega_1^2 \mu_s^2(0) \gamma z/\Omega^2_\mathrm{eff}$ in the IQZ
regime. For the same pump intensity, the latter is a factor of $ \gamma
z\gg 1$ greater than the former. (Here, $\gamma z\gg1$ is from $
\gamma \gg \Omega_1 {\mu^0_s}$ and $\Omega_1{\mu^0_s} z>1$.) In
terms of the SHG suppression, which determines the switching
contrast, for the same pump intensity we obtain
\begin{equation}
    \frac{\mathcal{S}_\mathrm{CQZ}}{\mathcal{S}_\mathrm{IQZ}}=\frac{\Omega^2_\mathrm{eff}}{ \gamma^2} \ll
    1,
\end{equation}
showing a much stronger suppression effect in the CQZ regime.

\subsection{Switching Performance}
\label{pa}

In the previous section, we have solved for the wave dynamics of light
propagation in a $\chi^{(2)}$ waveguide, where analytical
results are obtained for the three dynamical regimes of interests.
In this section we use these results to characterize the
switching performance. We focus on the CQZ regime with $\gamma=0$,
as it leads to the strongest SHG-suppression effect. We consider
nearly degenerate signal and pump waves and use the approximation
$n_d\approx n_p$.

We consider super-Gaussian input signal and pump waves in the form
\begin{equation}
\label{eqn22}
    \mu_{s,p}(t)\propto \exp\left(-\frac{t^{2m}}{2\sigma^{2m}}\right),
\end{equation}
where $m=0,1,2\ldots$ is the order-number for the super-Gaussian waves. $m=0$ corresponds to a
flat continuous wave (CW), $m=1$ corresponds to a Gaussian pulse, and
$m>1$ corresponds to super-Gaussian pulses with increasingly flatter top. In the following, we assume the pulse shapes to be the same for the signal and pump waves.

To characterize the switching performance, two important metrics are
normally used: the photon loss during switching and the switching
contrast. The switching loss is the fraction of energy missing
from the two outputs. In all of our switches, the mean energy loss
for CW operation is
\begin{equation}
\label{eqn24}
    \mathcal{L}=\frac{I_s}{I_p},
\end{equation}
with an upper-bound of $2\frac{I_s}{I_p}$. Here, $I_s$ ($I_p$)
is the input intensity of the signal (pump) wave. The universality
of $\mathcal{L}$ over all dynamical regimes has been discussed in
section \ref{wd}. For pulsed operation, where the signal and the pump have
similar shapes, $\mathcal{L}$ matches the CW result
(\ref{eqn24}). Thus, for all switches, a pump wave of power
over $100$ times greater than the signal wave is required to achieve below $1\%$ loss. In practice, this would fundamentally limit the present switch from being applied in a fan-in/fan-out all-optical network. This difficulty can be overcome with a Fabry-Perot design, which we will present later.

The switching contrast, on the other hand, is defined as the power
ratio with pump on and off in a certain output channel. As there
are two outputs in our switches, we define the ``flipped''
(``unflipped'') contrast as the power ratio in the flipped
(unflipped) output state. In Figs.~\ref{fig1} (a), (b), and~(c), the
flipped state respectively corresponds to the frequency state $2\omega_s$, the polarization state $|H\rangle-i|V\rangle$, and the output state in the ``q'' port. In each case, the unflipped state then corresponds to the complementary output.

\subsubsection{Frequency switching}
For CW operation, the flipped-state contrast is
\begin{equation}
\label{eqn20}
    \mathrm{FS}_\mathrm{CW}=\frac{2}{\mathcal{L}} \mathrm{tanh}^2\left(\frac{{\mu^0_s}\Omega_1
    L}{\sqrt{2}}\right),
\end{equation}
which increases monotonically with the waveguide length $L$. This
behavior arises from the fact that the SHG is a one-way process (assuming no spontaneous down-conversion). It approaches $2/\mathcal{L}$ when
$L\gg {\mu^0_s} \Omega_1$. The unflipped contrast, on the other
hand, is given by
\begin{equation}
    \mathrm{UF}_\mathrm{CW}=(1-\mathcal{L})\mathrm{cosh}^2\left(\frac{{\mu^0_s}\Omega_1 L}{\sqrt{2}}\right),
\end{equation}
which monotonically approaches infinity as $L\rightarrow \infty$ and all the power is transferred to the SH level.

In the case of pulsed operation with the pump off, the pulse
center up-converts more quickly than the pulse tails, due to the nonlinear nature of the SHG process.
After the switching, the SH pulse-width is reduced by half
compared to the signal. With the pump on, the SHG-dynamic is equally
suppressed across the whole pulse area. Therefore, the switching
contrasts with pump off or on are about the same as for CW inputs.

\subsubsection{Polarization and Spatial-Mode Switching}

The polarization and spatial-mode switches operate on a similar
principle; in fact, they yield an identical mathematical formulism.
Therefore, here we only analyze the spatial-mode switch, noting that all
results can be exactly applied to the polarization-mode switch.

In the case of CW inputs and small PNM (with $\Delta k\ll \Omega_1
{\mu^0_s}$), we set $L=2 z_0$ to necessarily acquire a $\pi$ phase
shift as the signal passes through the waveguide. We note that
$z_0$ is insensitive to $\Delta k$, but is proportional to
$1/{\mu^0_s}$; thus this switch can only function appropriately for
signal waves of a certain known intensity. Therefore, it can not be
applied to non-flat pulses. The flipped-state contrast in this
case is given by
\begin{eqnarray}
    \mathrm{FS}_\mathrm{CW}&=&\frac{16\Delta k^2}{\pi^2\Omega_1^4\mu^4_s(0)
    L^2 \mathcal{S}_\mathrm{CQZ}^2}>\frac{I_p^2}{I_s^2}.
\end{eqnarray}
This results in a high switching efficiency. The unflipped-state contrast,
on the other hand, is infinity in the ideal case, as no wave exits
from the ``p'' output port with the pump off.

In the case of large PNM with $\Delta k\gg \Omega_1^2 \mu_s^2(0)$,
the waveguide length is set to be
\begin{equation}
    L=\frac{2\pi \Delta k}{\Omega_1^2 \mu_s^2(0)} \gg 2z_0,
\end{equation}
to achieve a $\pi$ phase shift for the signal wave. Thus for a given pump,
a much longer waveguide is needed in the large PNM regime than
that in the small PNM regime. The flipped-state contrast in this
case is given by
\begin{equation}
\label{eqn23}
     \mathrm{FS}'_\mathrm{CW}=\frac{4}{\pi^2 \mathcal{S}_\mathrm{CQZ}^2}\approx \frac{4\Omega^4_\mathrm{eff}}{\pi^2\Delta k^4_1},
\end{equation}
which shows a moderate switching efficiency. For example, for $I_p=100
I_s$ and a typical $\Delta k=10 \Omega_1 {\mu^0_s}$, we have
$\mathrm{FS}'_\mathrm{CW}=40$. The unflipped-state switching
contrast here is infinity, for the same reason as in
the small PNM case.

\begin{figure}
\centering \epsfig{figure=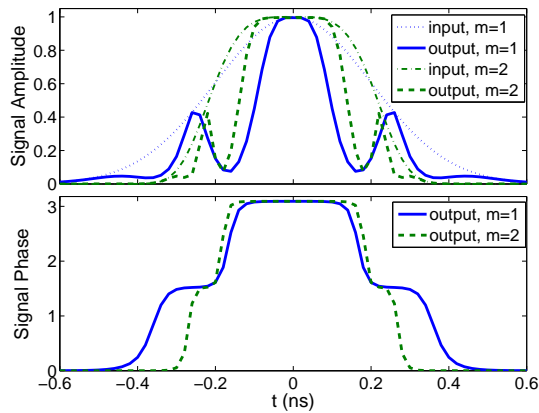, width=8.0cm} \caption{(Color online) Amplitude and phase profiles of the output signal wave passing
through a waveguide in the case of small PNM and pump off, shown
for Gaussian (m=1) and super-Gaussian (m=2) pulses. Here
$\sigma=0.2$ ns and all other relevant parameters are the same as
in Fig.~\ref{fig4}. \label{fig8}}
\end{figure}

When non-flat pulses are used, the system behavior is similar to
the CW case with the pump on. This is because the SHG-suppression
effect is determined by the ratio of the signal and pump
intensities, and not by their absolute values. Thus the flipped-state
contrast is about the same as in the CW case. On the other hand, with the pump off---due to the nonlinear nature of SHG---the output signal pulse will be significantly distorted. This gives rise to a low unflipped-state contrast.

First, for small PNM, the pulse distortion arises from the linear
dependence of the SHG period on the signal amplitude $\mu_s$. This is a fundamental difficulty inherent to SHG; to overcome this pulse distortion one would require nearly-flat-top super-Gaussian pulses. In Fig.~\ref{fig8} we show the amplitude and phase profiles of the output signal in the
fundamental level with the pump off. For a Gaussian pulse, the
amplitude is very distorted and the phase is highly
inhomogeneous. By using $m=2$ super-Gaussian pulses, the
distortion is somewhat mitigated. A direct consequence of the pulse
distortion is a low unflipped-state switching contrast. For the
parameters used in Fig.~\ref{fig8}, we numerically find the
contrast to be 10 for $m=1$; and the contrast near-linearly increases to $90$ for $m=10$.

\begin{figure}
\centering \epsfig{figure=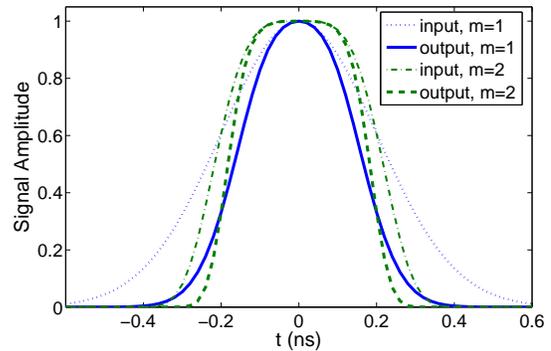, width=8.0cm} \caption{(Color online) Amplitude profile of the signal wave at the output port I for Gaussian (m=1)
and super-Gaussian inputs (m=2) with pump off in the case
of large PNM, compared to the input. Relevant parameters are the
same as in Fig.~\ref{fig5}. \label{fig10}}
\end{figure}

For large PNM, the signal pulse largely maintains
its shape with the pump off, while its phase evolves
inhomogeneously. For an $m$th-order Gaussian input, the resulting
phase shift is
\begin{equation}
    \phi_s=\pi \exp\left(-\frac{t^{2m}}{\sigma^{2m}}\right).
\end{equation}
This arises from the linear proportionality of the phase to the
intensity, as seen in Eq.~(\ref{eqn10}). As a result, the output
pulse in the ``p'' channel in Fig.~\ref{fig1}(c) is
\begin{equation}
   A^p_{s}\propto \exp\left[i\pi \exp\left(-\frac{t^{2m}}{\sigma^{2m}}\right)-1\right]  \exp\left(-\frac{t^{2m}}{2\sigma^{2m}}\right).
\end{equation}
Here, the first exponential term gives rise to pulse distortion.
As an example, we plot the signal output in Fig.~\ref{fig10} for Gaussian (m=1) and super-Gaussian (m=2) input pulses. Comparatively, the distortion is significantly mitigated
for super-Gaussian pulses. Lastly, for the parameters used in Fig.~\ref{fig5}, we numerically find the unflipped-state contrast to be
8 for $m=1$, and near-linearly increasing to $54$ for $m=10$.

\section{Waveguide Fabry-Perot Design}
\label{FPC}
The fundamental difficulty with the traveling-wave design is pulse distortion. To overcome this difficulty, in this section we present a Fabry-Perot design. The organization is as follows: In section \ref{FPC-setup}, we describe the switch setup and present the dynamical model. In section \ref{QA}, we use semi-static analysis to characterize the switching performance. Finally, in section \ref{pwm}, we present numerical simulations for pulsed inputs.

\subsection{The Model}
\label{FPC-setup}
\begin{figure}
\centering \epsfig{figure=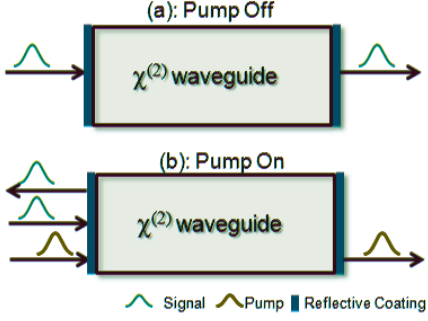, width=6.0cm} \caption{(Color online) Schematic setup of the noninteracting all-optical switch configured using a $\chi^{(2)}$ waveguide. Here (a) and (b) illustrate the ideal light paths with the pump off and on, respectively. \label{fig12}}
\end{figure}

The schematic setup of our Fabry-Perot switch is shown in Fig.~\ref{fig12}. It is composed of a $\chi^{(2)}$ waveguide designed to support PM or QPM SFG for the signal and pump waves. The two waveguide ends are coated with highly-reflective layers. The waveguide length is chosen such that both the signal and pump waves are in cavity resonance. With the pump off, as shown in Fig.~\ref{fig12}(a), the signal wave resonantly couples to the cavity, say, from the left, which then eventually exits from the right. For switching, we apply a pump pulse to change the cavity resonance condition via SFG, as shown in Fig.~\ref{fig12}(b). As a result the signal will be reflected (which is the switching operation). For optimal performance, the pump pulse is applied slightly ahead of the signal pulse to allow a sufficient pump field to built up in the cavity by the time the signal pulse arrives. This is necessary to reflect the front of the signal pulse. By excluding the signal wave from the cavity, the pump pulse is able to pass through the cavity unaffected. Note that while Fig.~\ref{fig12} shows a planar Fabry-Perot cavity, in practice a spherical cavity will be more robust against mirror divergence or misalignment. Of course, one can also replace the reflective coatings with two mirrors. It is straightforward to extend our analysis to these types of setups.

For simplicity, in the following we consider identical, lossless coatings on the two end faces. The coating reflectivity is $R$ (thus the transmissivity is $T=1-R$), which is assumed to be the same for both the signal and pump waves. For the SF field, it is taken
as $R'$ and in general $R'\neq R$. For further simplicity, we neglect the waveguide loss for both the signal and pump waves. This approximation is valid given that the single-round-trip loss is expected to be much smaller than the coating transmissivity. The loss rate for the SF wave, which determines whether the switch is in the IQZ or CQZ regime, is denoted by $2\gamma$.

To model the SFG process, we define the following dimensionless variables in a similar fashion as in Eq.~(\ref{eq-define}):
\begin{equation}
    A_j=i\sqrt{\frac{\epsilon_0 V n_j}{2\hbar\omega_j}} E_j,
\end{equation}
where $E_j$ ($j=s, p, f$) is the slowly-varying electric field of the signal, pump, and SF waves, respectively. We also introduce an effective Rabi-frequency similar to Eqs.~(\ref{eqn2}) and~(\ref{eqn3}),
\begin{eqnarray}
    \Omega &=& 2 d_{\mathrm{eff}} \sqrt{\frac{2\hbar\omega_s \omega_p \omega_f}{n_s n_p n_f \epsilon_0 c^2 V}}.
\end{eqnarray}

At phase matching, the SFG dynamics in the cavity are governed by
\begin{eqnarray}
\centering
\label{eqPFC1}
    \frac{\partial A_{s}}{\partial z}&=&-\Omega A^\ast_p A_{f}, \\
\label{eqPFC2}
    \frac{\partial A_{p}}{\partial z}&=&-\Omega A^\ast_s A_{f}, \\
\label{eqPFC3}
    \frac{\partial A_{f}}{\partial z}&=&\Omega A_{s} A_{p}-\gamma A_f.
\end{eqnarray}
The above equations of motion have analytical solutions under the
undepleted-pump approximation, in which $A_p(z)=A^0_p e^{i\phi_p(z)}$. In
the IQZ regime, corresponding to $\gamma\gg \Omega |A_p|$, we have
\begin{equation}
    A_s(z)\approx \exp\left[-\Omega_\mathrm{eff}^2 z/\gamma\right] A_s(0),
\end{equation}
where we have introduced $\Omega_\mathrm{eff}=\Omega A^0_p$.
In the opposite CQZ regime, we have
\begin{eqnarray}
    & & \!\!\! \!\!\!\!\!\!A_s(z)=A_s(0) \cos(\Omega_\mathrm{eff} z)-A_f(0) e^{-i\phi_p} \sin(\Omega_\mathrm{eff} z), ~~\\
    & &\!\!\!\!\!\!\!\!\! A_f(z)=A_f(0) \cos(\Omega_\mathrm{eff} z)+A_s(0) e^{i\phi_p} \sin(\Omega_\mathrm{eff} z).
\end{eqnarray}

\subsection{Quasi-Static Analysis}
\label{QA}
\begin{figure}
\centering \epsfig{figure=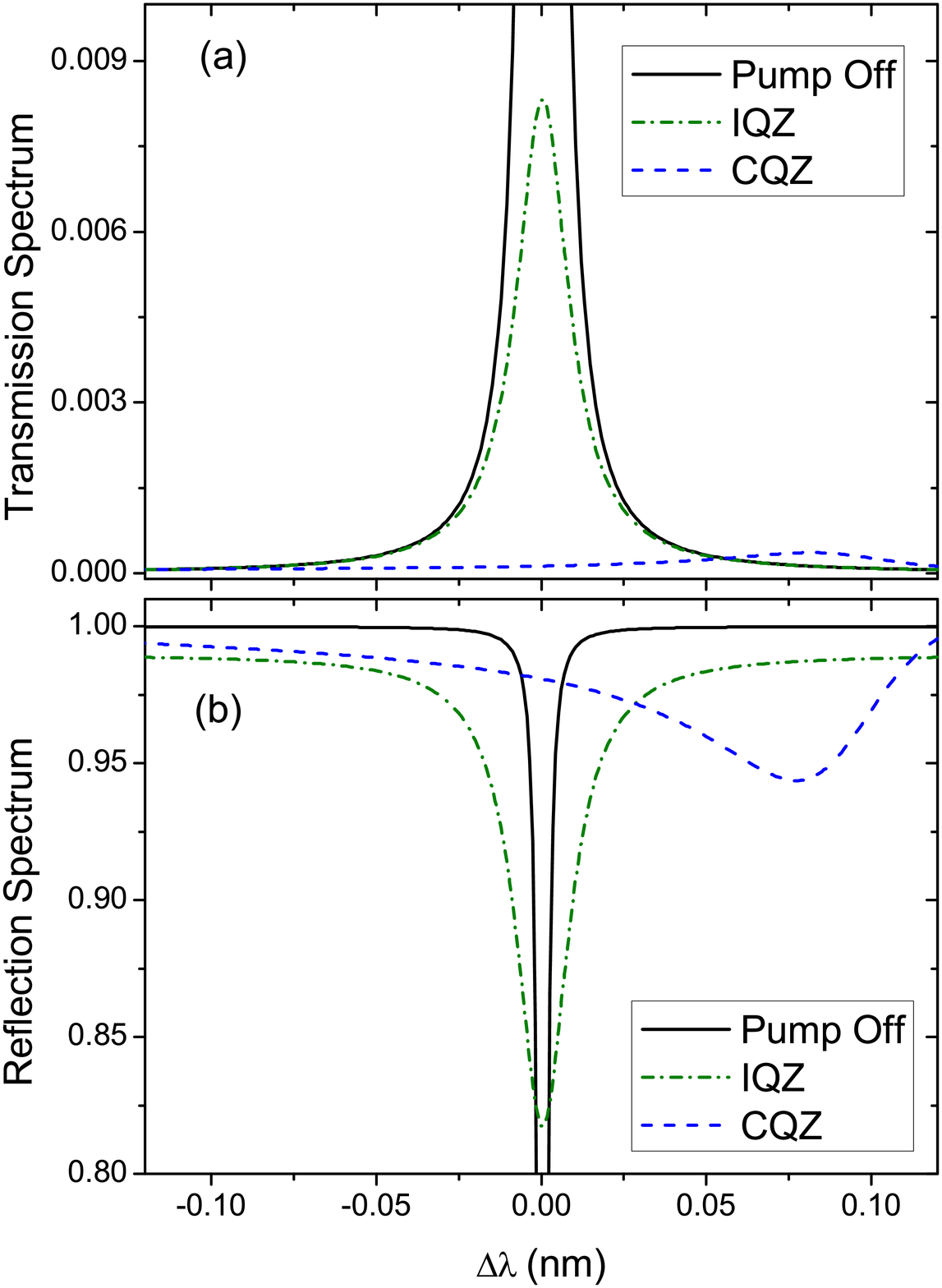, width=6.5cm} \caption{(Color online) Cavity transmission (a) and reflection (b) spectra, for the cases of pump off (solid), pump on in the IQZ regime (dash-dotted), and pump on in CQZ regime (dashed). The various parameters are chosen to be: $L=1$
mm, $R=0.99$, $R'=0.90$, and $\Omega_\mathrm{eff}=0.5$ mm$^{-1}$. The IQZ and CQZ regimes correspond to $\gamma=5$ mm$^{-1}$ and $\gamma=0$, respectively. \label{fig13}}
\end{figure}

We now use quasi-static analysis to characterize the switching performance. With the pump off, the signal wave is resonant with the cavity, as shown in Fig.~\ref{fig12}(a). Assuming a $\pi/2$ phase shift on the reflected light by the coating, a straightforward self-consistent analysis reveals
\begin{eqnarray}
      A^{T}_s &=&\frac{A^I_s T e^{i k_s L}}{1+R e^{2 i k_s L}},\\
      A^{R}_s &=& i \sqrt{R} A^I_s \left(1+\frac{T }{R+e^{-2 i k_s L}}\right),
\end{eqnarray}
where $k_j=n_j\omega_j/c$, and $A^T_s$ and $A^R_s$ are the transmitted and reflected signal fields, respectively. The cavity transmittance and reflectance spectra are then given by
\begin{eqnarray}
    T_\mathrm{off}(k_s) &=&\frac{(1-R)^2}{|1+R e^{i 2k_sL}|^2}, \\
    R_\mathrm{off}(k_s)&=& 1-\frac{(1-R)^2}{|1+R e^{i 2k_sL}|^2}.
\end{eqnarray}
An example of a typical cavity spectrum with the pump off is shown in Fig.~\ref{fig13}.

When the pump is on, the cavity spectrum is altered by the SFG process. In the IQZ regime with $\gamma\gg \Omega_\mathrm{eff}$, we find
\begin{eqnarray}
          A^{T}_s &=&\frac{A^I_s T e^{i k_s L}  e^{-\Omega^2_\mathrm{eff} L/\gamma}}{1+R e^{2 i k_s L} e^{-2\Omega_\mathrm{eff}^2 L/\gamma}},\\
      A^{R}_s &=& i \sqrt{R} A^I_s \left(1+\frac{T}{R+e^{-2 i k_s L}  e^{2\Omega_\mathrm{eff}^2 L/\gamma}}\right),
\end{eqnarray}
which lead to the following transmission and reflection spectra in the IQZ regime:
\begin{eqnarray}
    T_\mathrm{IQZ}(k_s) &=&\frac{(1-R)^2 e^{-2\Omega_\mathrm{eff}^2 L/\gamma}}{|1+R e^{i 2k_sL}e^{-2\Omega_\mathrm{eff}^2 L/\gamma}|^2}, \\
    R_\mathrm{IQZ}(k_s)&=& R^2 \left|\frac{1+e^{-2 i k_s L}  e^{2\Omega_\mathrm{eff}^2 L/\gamma}}{R+e^{-2 i k_s L}  e^{2\Omega_\mathrm{eff}^2 L/\gamma}}\right|^2.
\end{eqnarray}
On the signal resonance with $k_s L=(n+1/2)\pi$ ($n$ is an integer), they become
\begin{eqnarray}
    T^0_\mathrm{IQZ} &=&\frac{(1-R)^2 e^{-2\Omega_\mathrm{eff}^2 L/\gamma}}{|1-R e^{-2\Omega_\mathrm{eff}^2 L/\gamma}|^2}, \\
\label{eqnFPC-3}
    R^0_\mathrm{IQZ}&=& R^2 \left(\frac{1-e^{2\Omega_\mathrm{eff}^2 L/\gamma}}{R-e^{2\Omega_\mathrm{eff}^2 L/\gamma}}\right)^2.
\end{eqnarray}
Equation~(\ref{eqnFPC-3}) shows that the cavity reflectance $R^0_\mathrm{IQZ}$ decreases with $\gamma$, indicating that the switching is in fact less efficient in the presence of a strong loss for the SF wave.

In the CQZ regime with $\gamma\ll \Omega_\mathrm{eff}$, a similar quasi-static analysis can be performed. The cavity transmittance and reflectance at the signal resonance are approximately
\begin{eqnarray}
    T^0_\mathrm{CQZ} = \frac{T^2 (\eta^2+R'^2\zeta^2)}{\left|1-R \eta^2-i\sqrt{R'} \zeta [(i\sqrt{R}+\sqrt{R'})\eta+\sqrt{R}\zeta]\right|^2}, ~~\\
    R^0_\mathrm{CQZ} = \frac{(R+R')\zeta^2 (\sqrt{R R'} \eta+\zeta)^2}{\left|1-R \eta^2-i\sqrt{R'} \zeta [(i\sqrt{R}+\sqrt{R'})\eta+\sqrt{R}\zeta]\right|^2},~~
\end{eqnarray}
with $\eta=\cos(\Omega_\mathrm{eff} L)$ and $\zeta=\sin(\Omega_\mathrm{eff} L)$.

To show the switching performance, in Fig.~\ref{fig13} we plot the cavity spectra in three cases: pump off, pump on in the IQZ regime, and pump on in the CQZ regime. The system parameters are chosen to be $L=1$ mm, $R=0.99$, $R'=0.90$, and $\Omega_\mathrm{eff}=0.5$ mm$^{-1}$. In the IQZ and CQZ regimes, the dissipation of the SF wave is taken to be $\gamma=5$ mm$^{-1}$ and $\gamma=0$, respectively. As shown in Fig.~\ref{fig13}(a), the transmission window is significantly altered in both regimes. The cavity transmittance is lowered to $<1\%$ ($<0.05\%$) in the IQZ (CQZ) regime, while the corresponding reflectance is raised to $>81\%$ ($>98\%$) [see Fig.~\ref{fig13}(b)]. Here the fact that the transmittance and reflectance do not sum to 1 is indicative of signal losses. Comparatively, for similar parameters, the switch is more efficient in the CQZ regime than in the IQZ regime.

\begin{figure}
\centering \epsfig{figure=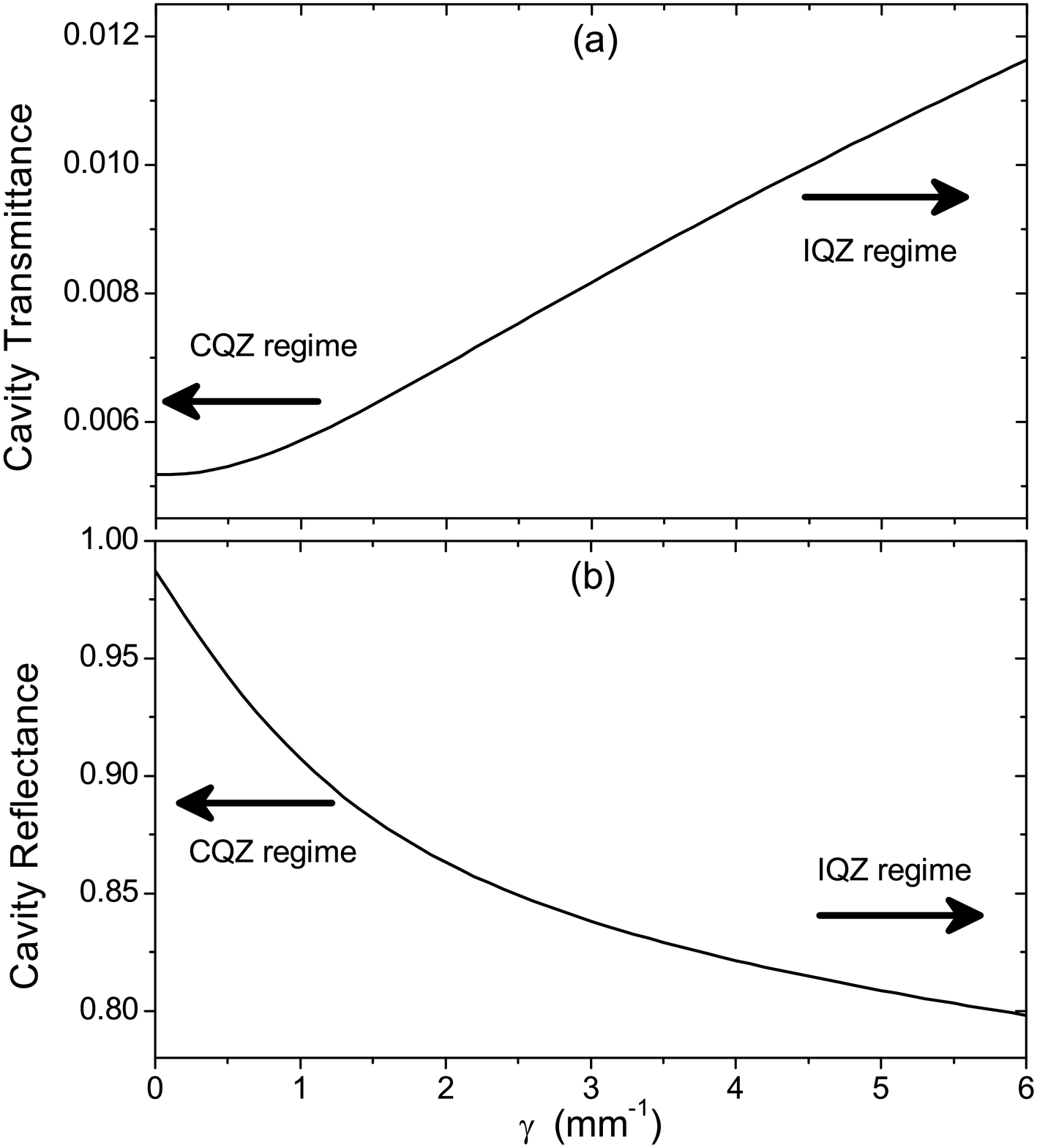, width=7.5cm} \caption{Cavity transmittance (a) and reflectance (b) plotted as a function of $\gamma$ at the signal resonance. The system parameters are the same as in Fig.~\ref{fig13}. \label{fig15}}
\end{figure}

To further compare the IQZ and CQZ effects, in Fig.~\ref{fig15} we plot the cavity transmittance and reflectance at the signal resonance as a function of $\gamma$. Unlike in Fig.~\ref{fig13} where we applied a static self-consistent analysis, here we obtain the results via an asymptotical analysis using a dynamical model. As shown in Fig.~\ref{fig15}(a), the cavity transmittance increases slowly
with $\gamma$, indicating that both the IQZ and CQZ mechanisms are effective in making the cavity opaque. In contrast, the reflectance drops dramatically from $99\%$ to $80\%$ when $\gamma$ increases from $0$ to $6$, as seen in Fig.~\ref{fig15}(b). This behavior shows that the cavity is highly reflective only in the CQZ regime. In the IQZ regime, on the other hand, a significant portion of the signal is lost owing to the presence of the SF-dissipation channel. It is, therefore, optimal to operate the switch in the CQZ regime.

We now give a physical explanation as to why all-optical switching is more efficient in the CQZ regime than in the IQZ regime. It is well known that light transmission through a cavity relies on the destructive interference of the reflected light and the light leaking from the cavity through the input coupler (the left-end coating in our design). Cavity reflectivity is then achieved by eliminating this destructive interference. There are two ways to accomplish this: (a) by diminishing the amplitude of the leaking light or (b) by modulating its phase. The JF proposal is based on the principle in (a), where the signal field in the resonator is eliminated via two-photon absorption \cite{JacFra09}. In this case, the cavity resonance is \emph{destroyed}. The switches in Ref. \cite{VanIbrRit02,IbrCaoKim03,TapLaiLan02, AlmBarPan04,Switch-Photonic-crytal-2005,Switching-PC2009,Switching-Photonic-Crystal-2005-2}, for example, are based on the principle in (b), where the cavity resonance is \emph{shifted} by modifying the refractive index of the intra-cavity material. Comparatively, the latter design allows a higher reflectance and weaker photon loss for similar parameters. This is because unlike in the JF switch, here a portion of the light entering the cavity is ``recycled'' as it leaks backwards from the cavity and eventually recombines with the reflected beam.

In our design, both the amplitude and phase modulation effects exist. In the IQZ regime, the dominant processes are a combination of the two and subsequent loss of the signal and the pump fields from the cavity. In this case, the switch operates on amplitude modulation, or more precisely, amplitude dissipation. This is exactly analogous to the mechanism in the JF switch, as both are based on two-photon absorption. The ultimate cavity reflectance is $R$, which is achieved when the
two-photon absorption is $100\%$. In the CQZ regime, on the other hand, the cavity fields undergo a coherent dynamics. Depending on the value of $R'$ (the reflectivity of the coatings for the SF light), the switch can be operated on either principle. In the first case, for $R'\ll 1$, the SF field is mostly lost through the coatings, resulting in a strong loss of the signal field from the cavity after each cavity cycle. The switch is then effectively a IQZ-based switch, working on amplitude modulation. Note, however, that in this case the depletion of the signal field in single pass through the cavity is efficient, as it is not suppressed by the Zeno effect. In contrast, in the JF switch (as well as in our switch in the IQZ regime), the two-photon absorption is Zeno-suppressed by the fast decay of their atomic level 3 (of the SF wave in our design) \cite{JacFra09}. In the second case, for $R'\approx 1$, the scattered SF light is mostly stored in the cavity. The signal field then undergoes a coherent, oscillatory dynamical evolution, incurring a $\pi/2$ phase shift. As a result, the destructive interference at the input coupler is destroyed, leading to reflectance from the cavity. The ultimate reflectance is $(1+R)/2$, which is higher than in the IQZ regime above by an amount of $(1-R)/2$. This improvement can be very useful for cavity-based all-optical switching, especially when short pulses are used \cite{Q-limit}. Working on phase-modulation, this switch is similar to the designs in Refs. \cite{VanIbrRit02,IbrCaoKim03,TapLaiLan02, AlmBarPan04,Switch-Photonic-crytal-2005,Switching-PC2009,Switching-Photonic-Crystal-2005-2}.
However, there is an important difference. In the cited designs, the pump photons are destroyed, for example, to create free carriers, leading to physical energy dissipation. This feature also prevents their potential operation in the quantum domain due to an increased number of noise channels that are coupled in. Our scheme, however, does not involve dissipation of any kind. In fact, \emph{it is the potential for the SFG process that changes the cavity from transmissive to reflective}. In this sense, our switch is very ``clean'' and, therefore, suitable for quantum applications (as we recently showed for an equivalent microdisk model \cite{HuangKumar10}).

\begin{figure}
\centering \epsfig{figure=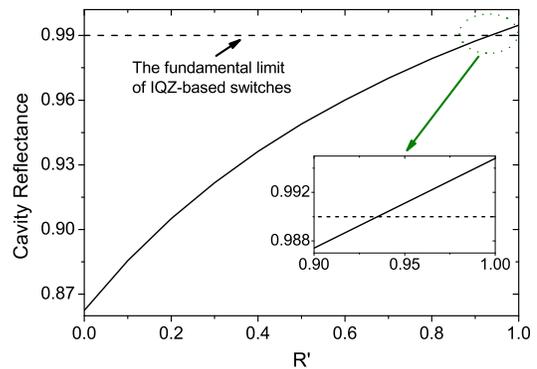, width=7.0cm} \caption{(Color online) Solid line:
Cavity reflectance as a function of $R'$, where $\gamma=0$ and all other
parameters are the same as in Fig.~\ref{fig15}. Dashed line: Fundamental limit for an IQZ-based switch. For the JF switch and the present
Fabry-Perot switch in the (effective) IQZ regime, this limit is obtained when the (effective) two-photon absorption is $100\%$. \label{fig19}}
\end{figure}

To graphically present these arguments, in Fig.~\ref{fig19} we plot the cavity reflectance as a function of $R'$. As shown, the cavity reflectance increases monotonically with $R'$. At $R'=0$, for which the cavity is effectively in the IQZ regime, the reflectance is only $86.3\%$. [Note that by increasing the pump power or the cavity length, this reflectance can eventually approach the limit of $99\% (=R)$ for an IQZ-based switch.] At $R'=1$, the reflectance increases to $99.5\%$; a result which agrees with our prediction above. From the inset
in Fig.~\ref{fig19}, one sees that for $R'>0.93$ the reflectance exceeds the fundamental limit ($99\%$) of IQZ-based switches, e.g., the JF switch and our switch in the (effective) IQZ-regime. We note that the crossing will occur at a smaller $R'$ as one increases the pump power.

\subsection{Pulse Dynamics}
\label{pwm}
In the last subsection we characterized the Fabry-Perot switch using a continuous-wave model. In this section we examine the switching performance for pulsed operation, focusing on the optimal CQZ regime with $R'=1$. We note first that for any cavity-based switch, the spectral width of the cavity must be much wider than the bandwidth of the incident pulse in order to avoid distortion of the switched output pulse \cite{Q-limit}. In our design, the spectral width of the cavity is
\begin{equation}
    \Delta_\mathrm{cavity}=\frac{v_c (1-R)}{L},
\end{equation}
where $v_c$ is the speed of light in cavity. For typical $v_c=10^{11}$ mm/s and $L=1$ mm, one would then require $R<0.99$ in order to switch a GHz-bandwidth pulse.

In order to reflect the entire signal pulse, a sufficient pump field must build up in the cavity by the time the signal pulse arrives. Therefore, when using signal and pump pulses of similar lengths, one must arrange to have the pump pulse arrive ahead of the signal pulse by a time interval $\tau_d$. Logically, $\tau_d$ should be chosen 1/2 of the cavity lifetime, i.e.,
\begin{equation}
    \tau_{d}=\frac{1}{2\Delta_\mathrm{cavity}}.
\end{equation}
Note that for similar light intensities the roles of the signal and the pump are completely interchangeable; in fact, in the opposite case where the signal pulse arrives prior to the pump, the pump pulse will be reflected, while transmitting the signal pulse. This behavior is due to the bistability of such switches \cite{JacFra09}.

\begin{figure}
\centering \epsfig{figure=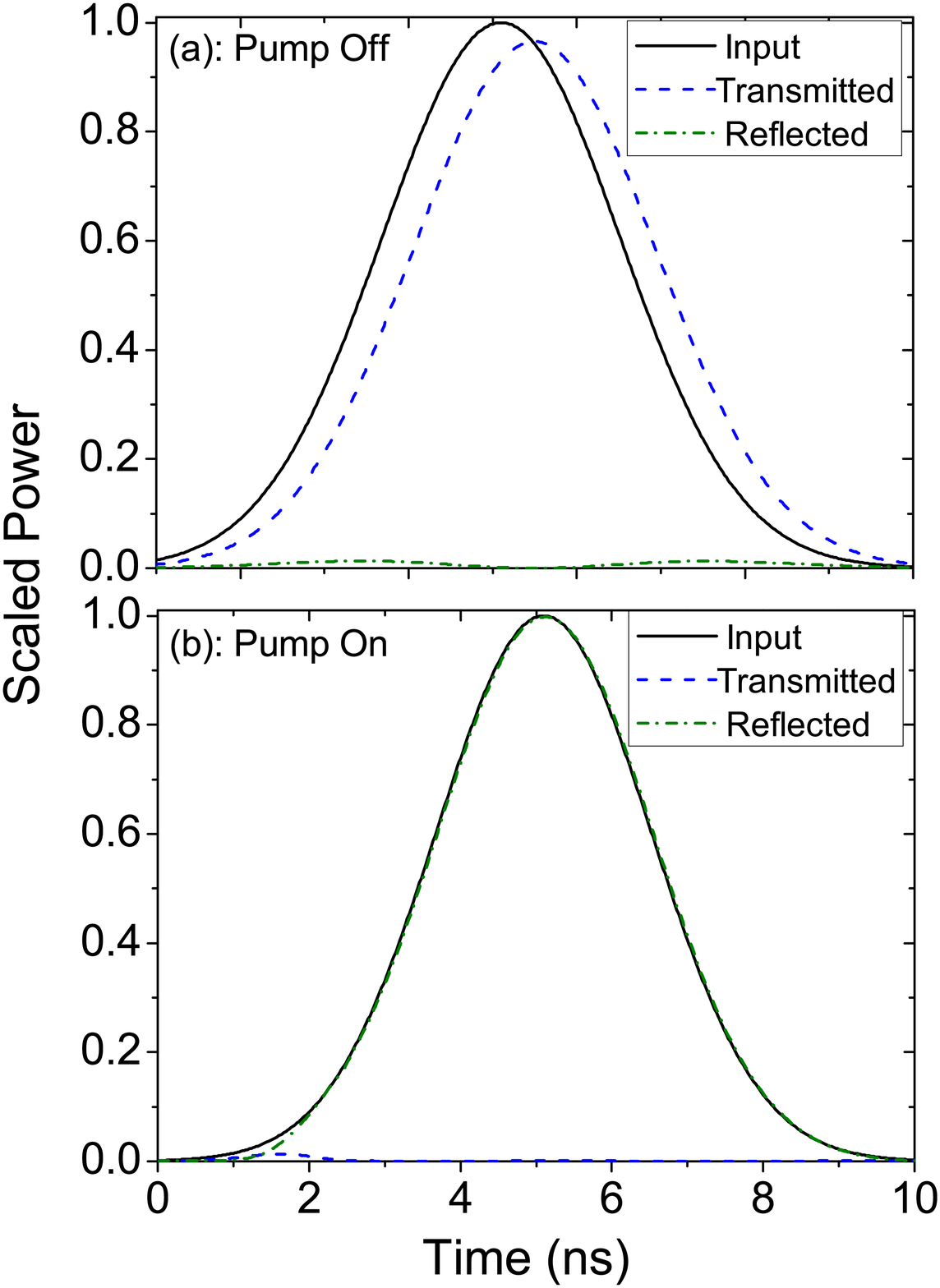, width=6cm} \caption{(Color online) Transmitted and reflected signal-pulse profiles with the pump off (a) and pump on (b). \label{fig16}}
\end{figure}

As an example, we consider $L=1$ mm, $R=0.95$, and $v_c=10^{11}$ mm/s. The signal and pump pulses are assumed to be identical, with a $e^{-1}$ half-width of $2$ ns. The peak intensity is chosen such that it gives $\Omega_\mathrm{eff}=0.5$ mm$^{-1}$. For these parameters, in Fig.~\ref{fig16} we show the temporal profiles of the transmitted and reflected signal pulses. As shown in Fig.~\ref{fig16}(a), with the pump off, the signal pulse is mostly transmitted, with a peak-to-peak time delay of $0.51$ ns and a transmitted power fraction of $98.1\%$. When the pump is applied, the signal pulse is mostly reflected (as high as $99.4\%$) as shown in Fig.~\ref{fig16}(b). These results are indicative of the high switching efficiency achievable with our Fabry-Perot switch design.

\begin{figure}
\centering \epsfig{figure=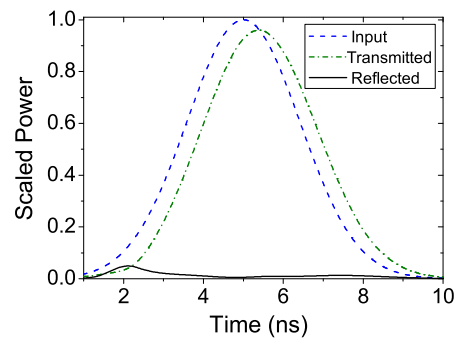, width=6.0cm} \caption{(Color online) (Color online) Pump-pulse profiles for the switching example plotted in Fig.~\ref{fig16}. \label{fig17}}
\end{figure}

In Fig.~\ref{fig17} we plot the profiles of the input, transmitted, and reflected pump pulses. As shown, the pump is mostly transmitted through the cavity with a well-maintained profile. The transmission and reflection fractions are $96.5\%$ and $3.2\%$, respectively, for the parameters used. The total pump-power loss is only $0.3\%$, indicating that this switch is indeed ``interaction-free''.

\section{Conclusion}
\label{dc}

In this paper, we have systematically studied a new, loss-free, scheme for all-optical switching that does not require direct interaction between the signal and pump waves, i.e, ``interaction-free'' switching. The distinct features of such switching devices include: a) photon dissipation via signal-pump coupling is eliminated; b) the photonic quantum states (both for the signal and the pump) are maximally protected from decoherence \cite{HuangKumar10}. We have presented a traveling-wave design and a Fabry-Perot design, both of which use $\chi^{(2)}$ waveguides for nonlinear interaction. After a thorough study of the switching parameter space, we conclude that the optimal switching performance is obtained in all cases in the coherent-quantum-Zeno (CQZ) regime. No photon dissipation occurs in this regime and the switching is achieved via the level splitting caused by the CQZ effect. The IQZ effect, which occurs when there is a strong dissipation channel, significantly degrades the switching efficiency, as well as increases the photon loss. These results are generally applicable to most interaction-free switching designs for optical applications.

\begin{acknowledgments}
This research was supported in part by the Defense Advanced
Research Projects Agency (DARPA) Zeno-based
Opto-Electronics program (grant W31P4Q-09-1-0014).
\end{acknowledgments}


\end{document}